# A Search Algorithm to Find Multiple Sets of One Dimensional Unipolar (Optical) Orthogonal Codes with Same Code-length and Low Weight

R. C. S. Chauhan, MIEEE, Y. N. Singh, SMIEEE, R. Asthana, MIEEE
Email: ram.hbti123@gmail.com

*Abstract*— **This paper describes a search algorithm to find multiple sets of one dimensional unipolar (optical) orthogonal codes characterized by parameter $(n, w, \lambda_a, \lambda_c)$. Here binary code sequence of length 'n' ('n' bits) and weight 'w' (number of bit '1's in the sequence) with $w << n$, as well as $\lambda_a$ and $\lambda_c$ are auto-correlation and cross-correlation constraint respectively for the codes within a set. For a given code length 'n' and code weight 'w', all possible difference sets, with auto-correlation constraints lying from 1 to w-1, can be designed with distinct code serial number. For given cross-correlation constraint $\lambda_c$ from 1 to w-1, multiple sets can be searched out of the codes with auto-correlation constraints less than or equal to given auto-correlation constraint $\lambda_a$ using proposed algorithm. The searched multiple sets can be sorted as having number of codes not less than the upper bound of the sets given by Johnson's bound. These one dimensional unipolar orthogonal codes have their application in incoherent optical code division multiple access systems.**

*Indexing Terms*—**Auto-correlation, Cross-correlation, One dimensional unipolar (optical) orthogonal codes (OOC).**

## I. INTRODUCTION

THE binary sequence of length 'n' ('n' bits) and weight 'w' (number of bit '1's) along-with all $n$-1 circular shifted versions represent the same code. For length 'n' and weight 'w', one or more than one unipolar orthogonal codes can be generated in the same orthogonal code set following correlation properties given in equation (i) & (ii) [5]. All the codes in one set are always orthogonal to each other but not necessarily orthogonal to the codes in other sets. The unipolar (optical) orthogonal code can be generated by putting a single optical pulse at every position of bit '1' and no pulses at the positions of bit '0's in the sequence. Suppose two code words $X$ and $Y$ are selected from the same orthogonal code set,

$X = (x_0, x_1, ..., x_{n-1})$, $Y = (y_0, y_1, ..., y_{n-1})$; $\forall (x_t, y_t) \in (0,1)$

For these codes

$$\lambda_a \geq \sum_{t=0}^{n-1} x_t x_{t \oplus \tau} \quad \text{as well as}$$

$$\lambda_a \geq \sum_{t=0}^{n-1} y_t y_{t \oplus \tau}, \quad \text{for } 0 < \tau \leq n\text{-}1. \quad \text{(i)}$$

Further,

$$\lambda_c \geq \sum_{t=0}^{n-1} x_t y_{t \oplus \tau} \text{ or } \lambda_c \geq \sum_{t=0}^{n-1} y_t x_{t \oplus \tau}, \text{ for } 0 \leq \tau \leq n-1. \quad \text{(ii)}$$

Here $t \oplus \tau$ implies $(t + \tau) \mod (n)$.

The auto-correlation of any code with un-shifted version of itself results in weight 'w' which is also auto-correlation peak.

$$\sum_{t=0}^{n-1} x_t x_t = w. \quad \text{(iii)}$$

The overlapping of one code with its shifted version can result in *0* to *w-1* bits. The maximum of these overlappings out of all pairs formed with its shifted versions is called auto-correlation constraint $\lambda_a$ and $1 \leq \lambda_a \leq w-1$. The overlapping of one code with other code and its shifted sequences result in *0* to *w-1* bits. The maximum of these overlappings out of all pairs formed with one code and other alongwith its shifted versions is called cross-correlation constraint $\lambda_c$ and $1 \leq \lambda_c \leq w-1$.

For $\lambda_a = \lambda_c = \lambda$ where $1 \leq \lambda \leq w-1$, The maximum number of unipolar (optical) orthogonal codes, Z, in one set is given by the following Johnson bounds [5- 7].
Johnson's bound A is,

$$Z(n, w, \lambda) \leq \left\lfloor \frac{1}{w} \left\lfloor \frac{n-1}{w-1} \ldots \ldots \left\lfloor \frac{n-\lambda}{w-\lambda} \right\rfloor \right\rfloor \right\rfloor = J_A(n, w, \lambda).$$

The Johnson's bound B holds only when $w^2 > n\lambda$, and is,

$$Z(n, w, \lambda) \leq Min\left(1, \left\lfloor \frac{w-\lambda}{w^2 - n\lambda} \right\rfloor\right) = J_B(n, w, \lambda);$$

The improved Johnson's Bound C for any integer k, $1 \leq k \leq \lambda - 1$; such that $(w-k)^2 > (n-k)(\lambda-k)$, is given as

$$Z(n, w, \lambda) \leq \left\lfloor \frac{1}{w} \left\lfloor \frac{n-1}{w-1} \ldots \ldots \left\lfloor \frac{n-(k-1)}{w-(k-1)} h \right\rfloor \right\rfloor \right\rfloor = J_C(n, w, \lambda);$$

Here $h = Min\left(n\text{-}k, \left\lfloor \frac{(n-k)(w-\lambda)}{(w-k)^2 - (n-k)(\lambda-k)} \right\rfloor\right);$

$\lfloor a \rfloor$ denotes the largest integer value less than '*a*'. In short, $J_A$, $J_B$, and $J_C$ denotes the Johnson bound A, B, and C respectively [5, 7].

In [8-19], many optical orthogonal code generation schemes have been proposed. Those schemes can design only one set of optical orthogonal codes corresponding to very much specific values of $(n, w, \lambda_a, \lambda_c)$ and do not find all possible sets.

Conventionally unipolar orthogonal codes are represented with its weighted positions [5,6], which is not an unique representation of the code because weighted position always changed with every circular shift of the code.

In [5,6,12], there is another representation for optical orthogonal codes proposed and discussed for $\lambda_a = \lambda_c = 1$. This proposed representation is a difference set containing difference of weighted positions in serial and circular order of the orthogonal code sequence. In these difference sets, there is no repetitive elements or no repetitive difference of positions so that auto-correlation constraint of code itself is minimum which is equal to one for unipolar orthogonal codes. The pairs of such difference sets from the set of codes has no common elements or no common difference of positions which make these with minimum cross correlation which is also equal to one for unipolar orthogonal codes. The reason for minimum auto-correlation and cross-correlation constraint equal to one is that at least one overlapping of bits '1' of orthogonal binary sequence is always exist with its among all shifted versions and with other code of pair along-with among all shifted versions of other code.

In this paper a standard difference of positions representation (DoPR) of one dimensional unipolar (optical) orthogonal codes is proposed and discussed. This proposed representation is an unique representation for the unipolar orthogonal codes with auto-correlation and cross-correlation constraint lying from 1 to *w-1*. With this representation the calculation of auto-correlation and cross-correlation constraints is also proposed which is easier than conventional method of calculation of correlation constraints [3].

This standard DoPR makes genaration of unipolar codes very easy for given code length n and weight w with auto-correlation constraints lying from 1 to w-1. The generated codes can be numbered serially. By using some search algorithms [1,2,4], multiple code sets can be searched out for given auto-correation and cross-correlation constraints. These search algorithms become complex while grouping the codes of desired correlation constraints. The complexity of such search algorithms can be reduced by following the proposed search algorithms and codes generation with improved procedure in this paper.

## II. ORTHOGONAL CODE DESIGN

A general scheme to design all the unipolar orthogonal code sets (n,w) is desirable. The sets may contain maximum or less than maximum number of orthogonal codes. All the codes in one set are always orthogonal with each other but not necessarily orthogonal with the codes of other sets. These code sets can be put to use as per the requirement. At a time one can select one of these multiple sets for use in the system.

### A. Example

The binary sequence of length '*n*' and weight '*w*' has *n-1* circular shifted sequences representing same orthogonal code. These shifted sequence has a common property that difference of positions (DoP) of any two nearest bit '1's in circular order, is always same. For *w* positions of bit '1's, there are w difference of positions' representation depending on which bit '1' is chosen as starting point. Consider a code 1101000 with code length *n*=7 and weight *w*=3. For this code, we have DoPs as given in Table -1.

| Shifted sequence | Positions of 1's | Difference of Positions |
|---|---|---|
| *1101000* | ( 0,1,3 ) | ( 1,2,4 ) |
| *1010001* | ( 0,2,6 ) | ( 2,4,1 ) |
| *0100011* | ( 1,5,6 ) | ( 4,1,2 ) |
| *1000110* | ( 0,4,5 ) | ( 4,1,2 ) |
| *0001101* | ( 3,4,6 ) | ( 1,2,4 ) |
| *0011010* | ( 2,3,5 ) | ( 1,2,4 ) |
| *0110100* | ( 1,2,4 ) | ( 1,2,4 ) |

*Table-1: DoP representation for Code 1101000.*

This table shows that the code 1101000 can be represented as (1,2,4) or by its circular shifted version i.e. (2,4,1) or (4,1,2). One can conclude the following for such representation.

1. There are w difference of positions in the code representations.

2. 0 (zero) is never present in difference of positions (DoP).

3. (1,2,4) can be used as a standard representation for the code 1101000, instead of other difference of positions like representation (2,4,1) or (4,1,2).

4. The last DoP in the standard representation will be given as n- (sum of rest other DoP elements).

5. If the DoP representation is such that the last DoP is always greater than or equal to the rest of DoPs. Then last DoP lies between $\left\lceil \dfrac{n}{w} \right\rceil$ and (n-w+1) in the standard representation. Here $\lceil a \rceil$ represents smallest integer value greater than a.

6. All the codes of the set (*n*=7, *w*=3) can be written in standard DoP representation as
{ (1,1,5), (1,2,4), (1,3,3), (2,1,4), (2,2,3) }.

7. When the last DoP is greater than the rest of DoPs in a code

then this is considered as standard DoP representation. For the cases, when in a code representation, largest DoP occurs more than once, the code sequence has multiple representation with last DoP as largest element. These multiple representation can be standardized by comparing them as follows

(i) All multiple DoP representations with last DoP element as largest element are compared for minimum value of first element. The code with minimum value of first element and largest value of last element is considered as standard DoP representation.

(ii) If first minimum valued element occurs in more than one multiple DoP representation at first positions, all these multiple DoP representations are compared for minimum value of second element among all second elements of these multiple DoP representations. The code with minimum value of second element is considered as standard DoP representation.

(iii) If second minimum valued element occurs in more than one multiple DoP representation at second positions, all these multiple DoP representations are compared for minimum value of third element among all third elements of these multiple DoP representations. The code with minimum value of third element is considered as standard DoP representation. This process is followed further till we find only one DoP representation. Similarly we can also have other methods to choose single standard DoP representation. The code design using DoP representation, every new code selected is not circular shifted version of earlier one. Also no two code should have same DoP to avoid repetition.

8. The DoP representation of code can be converted into representations showing positions of '1's or weighted position representation (WPR) [2].
(i) First position of '1' can be fixed as 0 or any position lying between 0 to n-1.
(ii) Second position is (first position plus first value from difference of positions representation) mod n.
(iii) Third position is (second position plus second value from DoP representation) mod n.
(iv) Similarly i$^{th}$ position is ((i-1)$^{th}$ position plus (i-1)$^{th}$ value from DoP representation) mod n.
If code in difference of positions representation is (1,1,5). It can be converted into positions of bit '1's representation as (0, 0+1, 0+1+1) = (0, 1, 2) = 1110000;

9. Auto-correlation constraint $\lambda_a$

The auto-correlation of any sequence with its circular shifted version can results in any number from 0 to $w$-1. The maximum allowed value for this measure is called auto-correlation constraint $\lambda_a$ which lies between 0 to $w$-1. The auto-correlation constraint of a set of more than one orthogonal codes is the maximum value of auto-correlation constraints among the orthogonal codes of that set.

The auto-correlation constraint of an orthogonal code can be calculated by converting standard DoP representation of code into Extended DoP Matrix representation. Suppose a code in standard DoP representation is given as $(a_1, a_2, a_3, ... a_w)$ such that $a_w \geq (a_1, a_2, ..., a_{w-1})$ for code length $n = (a_1 + a_2 + ... + a_w)$ and weight w.
The Extended DoP Matrix representation will be given by

$$\begin{bmatrix} a_1 & a_1+a_2 & ... & a_1+a_2+...+a_{w-1} \\ a_2 & a_2+a_3 & ... & a_2+a_3+...+a_w \\ a_3 & a_3+a_4 & ... & a_3+a_4+...+a_1 \\ ... & ... & ... & ... \\ a_w & a_w+a_1 & ... & a_w+a_1+...+a_{w-2} \end{bmatrix}.$$

One can observe that in each row of this matrix, the element value increase monotonically and no element repeats again. Each row of this matrix is compared with all the other rows. We observe how many element are same between the two rows being compared. We can choose the pair for which numbers of common elements are maximum. The $\lambda_a$, maximum auto-correlation with non-zero shift, will be this number + 1.

For the case when n = w.p, where p is an integer, there exist a code with DoP (p,p,...,upto w times) for which auto-correlation for shift of w, 2w, ... , (p-1)w will be w and for all other shifts it will be zero. Thus for this code $\lambda_a$=w. This case can be avoided by restricting n and w such that $n \neq w.p$, or n and w are co-prime with each other.

10. Cross-correlation constraint $\lambda_c$

The calculation of cross correlation constraint of any two codes is done by comparing Extended DoP Matrix representation of two codes. Each row of first matrix is compared with all the rows in second matrix. When row i in first matrix is compared with row j of second matrix, let $N_{ij}$ elements be same. Cross correlation constraint $\lambda_c$ is given by $\underset{\forall i,j}{Max}(N_{ij}) + 1$.

Each row when compared to another row needs at most w comparison so elements are monotonically increasing, and all w rows are compared against w rows in second matrix so that complexity becomes of the order of $O(w^3)$.

Consider the codes DoP (1,1,5) and DoP (1,2,4) in extended DoP representations as given in table-2.

| Code 1 (1,1,5) Extended DoP | Code 2 (1,2,4) Extended DoP | $N_{1j}$ | $\lambda_c$ |
|---|---|---|---|
| $\begin{bmatrix} 1 & 2 \\ 1 & 6 \\ 5 & 6 \end{bmatrix}$ | $\begin{bmatrix} 1 & 3 \\ 2 & 6 \\ 4 & 5 \end{bmatrix}$ | $N_{11} = 1$ <br> $N_{12} = 1$ <br> $N_{13} = 0$ | Max $N_{1j} = 1$ <br> $\lambda_c = 1+1=2$ |

*Table -2 Calculation of Cross-correlation Constraint $\lambda_c$*

We can see that $\lambda_c$ is 2 in table-2.

11. After calculating the auto and cross-correlation constraint, we can form a square matrix (NxN) (Fig.1.) showing the values of auto and cross correlation constraints of all the N codes of code set (n,w). The auto-correlation constraints lie at the diagonal and cross-correlation constraints at the non-diagonal positions. For i=j, $\lambda_{ij}$ represent to auto-correlation constraint of code#i, and for $i \neq j$, $\lambda_{ij}$ represent to cross-correlation constraint of code#i with code#j. One should note that $\lambda_{ij} = \lambda_{ji}$.

$$\begin{array}{c} code \\ \#1 \\ \#2 \\ ... \\ \#N \end{array} \begin{array}{cccc} \#1 & \#2 & ... & \#N \end{array} \\ \begin{bmatrix} \lambda_{11} & \lambda_{12} & ... & \lambda_{1N} \\ \lambda_{21} & \lambda_{22} & ... & \lambda_{2N} \\ ... & ... & ... & ... \\ \lambda_{N1} & \lambda_{N2} & ... & \lambda_{NN} \end{bmatrix}$$

*Fig.1: Correlation Matrix for N unipolar orthogonal codes.*

By observing the correlation matrix (NxN) we can form one or more than one set of orthogonal codes for a given value of $\lambda_a$ and $\lambda_c$ [1,2].

B. *Algorithm for generation of orthogonal codes' sets*

Step 1: Input code length 'n' and weight 'w', such that w < n and initialize first (w-1) DoP elements of the code i.e. $a_1, a_2,...,a_{w-1}$ to '1'. The $a_w$ will be given by $a_w = (n - (a_1 + a_2 + ... + a_{w-1}))$.

Step 2: Generate all the codes of set (n,w) in standard representation of Difference of Positions $(a_1, a_2, ..., a_{w-1}, a_w)$ with enumeration.

(i) $a_w \geq (a_1, a_2, ..., a_{w-1})$ and

(ii) $\left\lceil \dfrac{n}{w} \right\rceil \leq a_w \leq (n - w + 1)$.

All the codes generated with condition $a_w > (a_1, a_2, ..., a_{w-1})$ will always be in standard DoP representation. While for the condition when $a_w$ is equal to any one or more than one of $(a_1, a_2, ..., a_{w-1})$, and greater than remaining DoP elements, the code has more than one representations as $a_w \geq a_1, a_2, ..., a_{w-1}$. In this condition, that representation is chosen for which

(i) $a_1$ is minimum, and

(ii) If minimum $a_1$ is found in more than one DoP representations, then minimum $a_2$ is searched among DoPs with minimum $a_1$. The DoP representation with minimum $a_1$ and minimum $a_2$ is considered as standard DoP representation. Similarly, search upto $a_{w-1}$ to find standard DoP representation may be needed if $a_1, a_2, ..., a_{w-2}$ are same in more than two members of candidate codes.

Step 3: Calculation of auto-correlation constraint $\lambda_a$

For each of the generated codes in step 3, the extended DoP matrix is formed, in this matrix of size w x (w-1), all the rows are compared with each other. The two rows which has maximum common elements, are chosen. The $\lambda_a$ will be maximum common number of elements of these two rows, plus one. The $\lambda_a$ for all generated codes is calculated similarly.

Step 4: Calculation of cross-correlation constraint $\lambda_c$

In Extended DoP matrix representation of any two codes, Each row of the EDoP of first code is compared with all the rows of EDoP of second code. The row in other code, which has maximum common elements, is selected. The maximum cross-correlation between the two codes will be number of common elements of these two rows plus one. The maximum cross-correlations for all other pair of codes is calculated similarly.

Step 5: Formation of correlation matrix

In step 2, let the number of generated code be N. A $N \times N$ matrix can be formed in such a way that it contains correlation of code x with code y, for $1 \leq (x, y) \leq N$.

When $x = y$, it represent maximum auto-correlation for non zero shift which form diagonal elements of $N \times N$ correlation matrix.

For $x \neq y$, the maximum cross correlation constraint of code#x with code#y is found as a non-diagonal element in row x and column y.

Step 6: Formation of sets of orthogonal codes for given values of $\lambda_a$ and $\lambda_c$.

First, all those codes are selected for which diagonal entries are $\leq \lambda_a$. All the rows and column for the codes not selected are removed from the matrix, giving a reduced correlation matrix. Within these codes, only those sets of codes are selected which has cross-correlation entries $\leq \lambda_c$ in reduced correlation matrix. For $\lambda_a = \lambda_c = (w-1)$, the maximum value of auto with non-zero shift and cross correlation, only one set of orthogonal codes will be achieved which will have all the codes generated in step 2 [1].

### III. COMPARISON WITH OTHER SCHEMES

a) One of the known method for generating OOC is based on prime sequences [8]. In this, a set of optical orthogonal codes (n,w, $\lambda_a$, $\lambda_c$) is generated for any prime number p. For these codes, the weight w = p, length n = $p^2$, auto-correlation constraint $\lambda_a$ =p-1 and cross-correlation constraint $\lambda_c$=2. The number of optical orthogonal codes in this set, are given by N=p.

b) In another type of OOC based on "Quasi Prime" [9], in code set $(n, w, \lambda_a, \lambda_c)$, $n = qp$, where $(r-1)p < q < rp$. Here p is a prime number; $q$ and $r$ are positive integers;
weight $w = q$,
auto-correlation constraint $\lambda_a = (p-1)r$,
cross-correlation constraint $\lambda_c = 2r$,
and the number of code-words $N = p$.

c) In OOCs based on Quadratic Congruences [10, 11], the orthogonal code set $(n, w, \lambda_a, \lambda_c)$ is constructed for the length $n = p^2$, weight $w = p$, $\lambda_a = 2$, and $\lambda_c = 4$. When the length of the code is extended, we get Extended Quadratic Congruence code. It can be used for construction of codes of length $n = p(2p-1)$, weight $w = p$, autocorrelation constraint $\lambda_a$ =1, and cross correlation constraint $\lambda_c = 2$.

d) In Projective Geometry based OOCs [12,13], a Projective Geometry $PG(m,q)$ of order m is constructed from a vector space $V(m+1,q)$ of dimension m+1 over $GF(q)$. Where $GF(q)$ is Galois Field with q elements. An s-space in a $PG(m,q)$ corresponds to an (s+1) dimensional space through the origin in $V(m+1,q)$ [11]. Here one-dimensional subspaces of V are the points and the two dimensional subspaces of V are the lines.

Number of points n in $PG(m,q)$, n= $\left(\dfrac{q^{m+1}-1}{q-1}\right)$ will give the length of the codeword. Number of points w= $\left(\dfrac{q^{s+1}-1}{q-1}\right)$ in the s-space, will give the weight of the codeword. The intersection of two s-spaces is an (s-1)-space.

Number of points in the (s-1)-space are $\lambda = \left(\dfrac{q^s-1}{q-1}\right)$ = max ($\lambda_a$, $\lambda_c$). The circular shift of an s-space is also an s-space. The orbit is the set of all s-spaces that are circular shift of each other. The number of code words is always equal to number of complete orbits. A codeword consists of discrete logarithm of points in each representative s-space. Total number of s-spaces is $M_s = \binom{n}{s+1} / \binom{w}{s+1}$.

Total number of code-words constructed using $PG(m,q)$ for given value of s are equal to $M = \left\lfloor \dfrac{M_s}{n} \right\rfloor$.

e) In OOCs based on Error Correcting Codes [14-17], a 't' error correcting code is represented by (n,d,w), where $n$ is length, $d$ is minimum hamming distance between any two code words, w is the constant weight of a code from the code-set. The minimum distance $d \geq 2t+1$. An OOC $(n, w, \lambda_a, \lambda_c)$ is equivalent to constant weight error correcting codes with minimum distance $d = 2w - 2\lambda$, where $\lambda$ is maximum of ($\lambda_a, \lambda_c$) [13]. Only those error correcting codes are selected for optical orthogonal code set whose circular shifted versions are also error correcting code word.

f) In Optical orthogonal codes using Hadamard Matrix, any Hadamard matrix of order n can be used to generate the matrix of order n-1 by deleting 1st row and 1st column. The rows of the matrix of order n – 1 form a code set. From the same code set, the repeated or circularally shifted codes are included as a single code to form the optical orthogonal code set of length $n$ =4$t$-1. Here weight w =2$t$-1; $\lambda_a$ = $t$-1; and $\lambda_c = t$; where $t$ is any positive integer [18].

g) In OOCs based on Balanced Incompleted Block Design (BIBD) [19], two families of OOC sets are constructed. First is (n, w, 1, 1) with optimal cardinality N, the second is (n, w, 1, 2) with cardinality 2N.

a1) (n, w, 1, 1) OOC for odd w
The weight $w = 2m+1$ will be odd for a positive integer m. The code length is given by $n = w(w-1)t+1$ where $t$ is chosen such that $n$ is a prime number.

a2) (n, w, 1, 1) OOC for even w
The weight $w = 2m$ will be even for a positive integer m. The code length is given by $n = w(w-1)t+1$ where $t$ is chosen such that $n$ is a prime number.

Here N is less than Johnson bound A for number of codes for (*n*, *w*, 1, 1). Similarly 2N is less than Johnson bound A for number of codes for (*n*, *w*, 1, 2).

The algorithm given in this paper find set of optical orthogonal codes (n, w, $\lambda_a, \lambda_c$) for weight w and length n, such that n and w are co-prime with each other. The auto-correlation constraint $\lambda_a$ and cross-correlation constraint $\lambda_c$ may take any values from 1 to w-1. The number of optical orthogonal codes in the set is given by N=J$_A$, the Johnson bound A, upper bound on number of codes in a set.

In above mentioned schemes (a to g), for (*n*, *w*, $\lambda_a, \lambda_c$), only one set of orthogonal codes is generated, but the algorithm proposed in this paper is able to generate all possible sets of orthogonal codes for (*n*, *w*, $\lambda_a, \lambda_c$).

We have verified our algorithm by implementing it and checking the orthogonality condition among the generated codes. We could also get the largest set of orthogonal codes as specified by Johnsons bounds. We not only get a single set of orthogonal codes with maximum cardinality but all possible sets of orthogonal codes with maximum cardinality. We could also generate code set for all values of n and w with only one condition of n and w being co-prime. Thus the method turns out to be much less restrictive in choice of n and w.

Generally in other schemes given in literatures [8-19], only one set of orthogonal codes with maximum or less than maximum cardinality can be achieved. A sample code set generated by presented algorithm for code length n =19, code weight w = 3, $\lambda_a = 1$, and $\lambda_c = 1$ is shown in Appendix. The drawback of proposed algorithm is its higher computational complexity $O\left(\left(\frac{n}{w}\right)^{w-1}\right)$ in comparison to other schemes proposed in literature [8-19]. These schemes have computational complexity of the order $O(n^2)$ for OOCs based on Hadamard Matrix [18], or $O(w^3)$ for OOCs based on Prime sequence [8], and OOCs based on Quadratic congruence [10]. The higher computational complexity of proposed algorithm may be lowered for w<<n, which is also the practical condition assumed in generation of unipolar (optical) orthogonal codes.

IV. CONCLUSION

This paper presents a general algorithm to design all possible unipolar codes of length 'n' and weight 'w' and their all possible orthogonal codes' sets. It will be very helpful in generating optical orthogonal codes for use in optical domain as well as in other fields where orthogonal codes are required.

This algorithm needs to be extended to two dimensional and three dimensional orthogonal code designs. It will be worth interestingly gain achieved by this algorithm for multi-dimensional codes. The computational complexity of the algorithm is high. One need to develop better algorithm with lower computational complexity. The biggest advantage offered over other schemes is generation of all possible sets of unipolar (optical) orthogonal codes.

## Appendix

The simulation result for program written as per algorithm given in this paper, for code length n=19, code weight w=3, $\lambda_a$ =1, and $\lambda_c$ =1 and Johnson bound $J_A$ =3

Input code length (n):19
Input weight (w):3

Codes formed = 51

| Code No. | DOPR | | | WPR | | | $\lambda_a$ |
|---|---|---|---|---|---|---|---|
| 1 | 1 | 1 | 17 | 0 | 1 | 2 | 2 |
| 2 | 1 | 2 | 16 | 0 | 1 | 3 | 1 |
| 3 | 1 | 3 | 15 | 0 | 1 | 4 | 1 |
| 4 | 1 | 4 | 14 | 0 | 1 | 5 | 1 |
| 5 | 1 | 5 | 13 | 0 | 1 | 6 | 1 |
| 6 | 1 | 6 | 12 | 0 | 1 | 7 | 1 |
| 7 | 1 | 7 | 11 | 0 | 1 | 8 | 1 |
| 8 | 1 | 8 | 10 | 0 | 1 | 9 | 1 |
| 9 | 1 | 9 | 9 | 0 | 1 | 10 | 2 |
| 10 | 2 | 1 | 16 | 0 | 2 | 3 | 1 |
| 11 | 2 | 2 | 15 | 0 | 2 | 4 | 2 |
| 12 | 2 | 3 | 14 | 0 | 2 | 5 | 1 |
| 13 | 2 | 4 | 13 | 0 | 2 | 6 | 1 |
| 14 | 2 | 5 | 12 | 0 | 2 | 7 | 1 |
| 15 | 2 | 6 | 11 | 0 | 2 | 8 | 1 |
| 16 | 2 | 7 | 10 | 0 | 2 | 9 | 1 |
| 17 | 2 | 8 | 9 | 0 | 2 | 10 | 1 |
| 18 | 3 | 1 | 15 | 0 | 3 | 4 | 1 |
| 19 | 3 | 2 | 14 | 0 | 3 | 5 | 1 |
| 20 | 3 | 3 | 13 | 0 | 3 | 6 | 2 |
| 21 | 3 | 4 | 12 | 0 | 3 | 7 | 1 |
| 22 | 3 | 5 | 11 | 0 | 3 | 8 | 1 |
| 23 | 3 | 6 | 10 | 0 | 3 | 9 | 1 |
| 24 | 3 | 7 | 9 | 0 | 3 | 10 | 1 |
| 25 | 3 | 8 | 8 | 0 | 3 | 11 | 2 |
| 26 | 4 | 1 | 14 | 0 | 4 | 5 | 1 |
| 27 | 4 | 2 | 13 | 0 | 4 | 6 | 1 |
| 28 | 4 | 3 | 12 | 0 | 4 | 7 | 1 |
| 29 | 4 | 4 | 11 | 0 | 4 | 8 | 2 |
| 30 | 4 | 5 | 10 | 0 | 4 | 9 | 1 |
| 31 | 4 | 6 | 9 | 0 | 4 | 10 | 1 |
| 32 | 4 | 7 | 8 | 0 | 4 | 11 | 1 |
| 33 | 5 | 1 | 13 | 0 | 5 | 6 | 1 |
| 34 | 5 | 2 | 12 | 0 | 5 | 7 | 1 |
| 35 | 5 | 3 | 11 | 0 | 5 | 8 | 1 |
| 36 | 5 | 4 | 10 | 0 | 5 | 9 | 1 |
| 37 | 5 | 5 | 9 | 0 | 5 | 10 | 2 |
| 38 | 5 | 6 | 8 | 0 | 5 | 11 | 1 |
| 39 | 5 | 7 | 7 | 0 | 5 | 12 | 2 |
| 40 | 6 | 1 | 12 | 0 | 6 | 7 | 1 |
| 41 | 6 | 2 | 11 | 0 | 6 | 8 | 1 |
| 42 | 6 | 3 | 10 | 0 | 6 | 9 | 1 |
| 43 | 6 | 4 | 9 | 0 | 6 | 10 | 1 |
| 44 | 6 | 5 | 8 | 0 | 6 | 11 | 1 |
| 45 | 6 | 6 | 7 | 0 | 6 | 12 | 2 |
| 46 | 7 | 1 | 11 | 0 | 7 | 8 | 1 |
| 47 | 7 | 2 | 10 | 0 | 7 | 9 | 1 |
| 48 | 7 | 3 | 9 | 0 | 7 | 10 | 1 |
| 49 | 7 | 4 | 8 | 0 | 7 | 11 | 1 |
| 50 | 8 | 1 | 10 | 0 | 8 | 9 | 1 |
| 51 | 8 | 2 | 9 | 0 | 8 | 10 | 1 |

available values of $\lambda_a$ were 1 and 2

For $\lambda_a$ =1, and $\lambda_c$ =1

Code Numbers which can be used for forming sets for $\lambda_a$ =1, and $\lambda_c$ =1 are

2   3   4   5   6   7   8   10   12   13   14   15   16
17   18   19   21   22   23   24   26   27   28   30   31
32   33   34   35   36   38   40   41   42   43   44   46
47   48   49   50

if $\lambda_a = \lambda_c$ =1, maximum codes in a set given by Johnson bound $J_A$ =3
Number of maximum code in a every set = 3.
The code sets are formed as following, given with their code numbers
*************** Code 3 ***************
   3   16   38
   3   16   44
   3   38   47
   3   44   47
If we begin with Code No. 3, No. of sets = 4.
*************** Code 4 ***************
   4   15   24
   4   15   48
   4   24   41
   4   41   48
If we begin with Code No. 4, No. of sets = 4.
*************** Code 5 ***************
   5   17   21
   5   17   28
   5   21   51
   5   28   51
If we begin with Code No. 5, No. of sets = 4.
*************** Code 7 ***************
   7   12   31
   7   12   43
   7   19   31
   7   19   43
If we begin with Code No. 7, No. of sets = 4.
*************** Code 12 ***************
  12   31   46
  12   43   46
If we begin with Code No. 12 and leaving the sets already mentioned with code No 12, No. of sets = 2.
*************** Code 15 ***************

```
15  24  26
15  26  48
```
If we begin with Code No. 15 and leaving the sets already mentioned with code No 15, No. of sets = 2.

*************** Code 16 ***************
```
16  18  38
16  18  44
```
If we begin with Code No. 16 and leaving the sets already mentioned with code No 16, No. of sets = 2.

*************** Code 17 ***************
```
17  21  33
17  28  33
```
If we begin with Code No. 17 and leaving the sets already mentioned with code No 17, No. of sets = 2.

*************** Code 18 ***************
```
18  38  47
18  44  47
```
If we begin with Code No. 18 and leaving the sets already mentioned with code No 18, No. of sets = 2.

*************** Code 19 ***************
```
19  31  46
19  43  46
```
If we begin with Code No. 19 and leaving the sets already mentioned with code No 19, No. of sets = 2.

*************** Code 21 ***************
```
    21  33  51
```
If we begin with Code No. 21 and leaving the sets already mentioned with code No 21, No. of sets = 1.

*************** Code 24 ***************
```
    24  26  41
```
If we begin with Code No. 24 and leaving the sets already mentioned with code No 24, No. of sets = 1.

*************** Code 26 ***************
```
  26  41  48
```
If we begin with Code No. 26 and leaving the sets already mentioned with code No 26, No. of sets = 1.

*************** Code 28 ***************
```
    28  33  51
```
If we begin with Code No. 28 and leaving the sets already mentioned with code No 28, No. of sets = 1.

All other remaining codes with $\lambda_a = 1$, do not form any more sets containing 3 codes in the set with $\lambda_c = 1$.

For code length n=19, code weight w=3, auto-correlation constraint $\lambda_a = 1$ and cross-correlation constraint $\lambda_c = 1$, with maximum 3 codes in each set, total Possible sets are 32, From all these 32 sets formed any one set can be selected for unipolar (optical) orthogonal codes. The codes of any one set are given in DOP representation which can be converted into orthogonal binary sequences using the procedure explained in the paper. These unipolar (optical) orthogonal codes can be assigned to multiple users of Incoherent Optical CDMA systems.

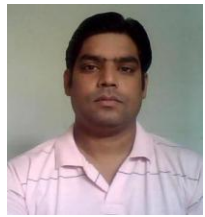

**Ram Chandra Singh Chauhan** received his B.Tech degree in Electronics Engineering from Institute of Engineering & Technology, Lucknow, University of Lucknow, (U.P.), India in 2001. He received his M.Tech degree in Digital System (Electronics Engineering) from Motilal Nehru National Institute of Technology, Allahabad, (U.P.), India, in 2003. He had worked as a Lecturer in Electronics & Communication Engineering department of University Institute of Engineering & Technology, Kanpur from January 2004 to July 2007. He received a Teacher Fellowship by Uttar Pradesh Technical University, Lucknow (U.P.) in July 2007 along-with pursue his PhD degree. He is presently working as Teacher Fellow (Lecturer) in Electronics Engineering department of H.B.T.I. Kanpur, India since July 2007 in parallel with pursuing his PhD degree from U.P.T.U. Lucknow. His interests are in Multiple Access Schemes for Optical Channel, Orthogonal coding theory, Digital Signal Processing and Information Theory and Coding.

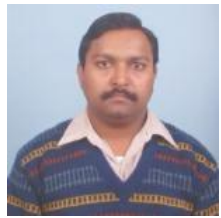

**Yatindra Nath Singh**, received his M.Tech degree in Optoelectronics and optical Communications from IIT Delhi in 1992. He received his PhD degree in Optical Communication Networks from Electrical Engineering department of IIT Delhi in 1997. Currently he is working with Electrical Engineering Department of IIT Kanpur as Professor.

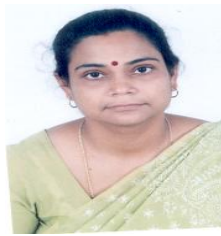

**Rachana Asthana**, received her M.Tech degree in Electronics & Communication from University of Roorkee ( now Indian Institute of Technology, Roorkee) in 1990. She received her Ph.D. degree (titled- Study of P-cycle based protection in optical networks and removal of its shortcomings) from Electrical Engineering department of Indian Institute of Technology, Kanpur in 2008. She is working as Associate Professor in Electronics Engineering department of H.B.T.I. Kanpur.